\documentclass[letterpaper, 10 pt, conference]{ieeeconf}  

\IEEEoverridecommandlockouts                              

\title{\LARGE \bf
Deep Learning for Visual Recognition of Environmental Enteropathy and Celiac Disease
}

\author{Aman Shrivastava$^{1}{^{\S}}$ Karan Kant$^{1}{^{\S}}$ Saurav Sengupta$^{1}{^{\S}}$ Sung-Jun Kang$^{1}{^{\S}}$%
\\ Marium Khan$^{2}$ 
S. Asad Ali$^{3}$ Sean R. Moore$^{2}$ Beatrice C. Amadi$^{4}$ Paul Kelly$^{4,5}$%
\\Donald E. Brown${^{1,6}{^*}}$  Sana Syed${^{2}{^*}}$ 
\thanks{$^{1}$Data Science Institute,
        University of Virginia, Charlottesville, VA 22903, USA
        {\tt\small}}%
\thanks{$^{2}$Department of Pediatrics, University of Virginia, Charlottesville, Virginia,USA
        {\tt\small}}%
\thanks{$^{3}$Department of Pediatrics \& Child Health, Aga Khan University,
Karachi, Pakistan
        {\tt\small}}%
\thanks{$^{4}$Tropical Gastroenterology and Nutrition group, University of Zambia School of Medicine, Lusaka, Zambia
        {\tt\small}}%
\thanks{$^{5}$Blizard Institute, Barts and The London School of Medicine, Queen
Mary University of London, London, United Kingdom
        {\tt\small}}%
\thanks{$^{6}$Department of Systems and Information Engineering, University of
Virginia, Charlottesville, VA 22904, USA 
        {\tt\small}}%
\thanks{$^{}$*Co-corresponding authors}%
\thanks{$^{\S}$Equal contribution}%
}

\usepackage{graphicx}
\graphicspath{ {./images/} }

\begin{document}

\maketitle
\thispagestyle{empty}
\pagestyle{empty}

\begin{abstract}

Physicians use biopsies to distinguish between different but histologically similar enteropathies. The range of syndromes and pathologies that could cause different gastrointestinal conditions makes this a difficult problem. Recently, deep learning has been used successfully in helping diagnose cancerous tissues in histopathological images. These successes motivated the research presented in this paper, which describes a deep learning approach that distinguishes between Celiac Disease (CD) and Environmental Enteropathy (EE) and normal tissue from digitized duodenal biopsies. Experimental results show accuracies of over 90\% for this approach. We also look into interpreting the neural network model using Gradient-weighted Class Activation Mappings and filter activations on input images to understand the visual explanations for the decisions made by the model.

\end{abstract}

\section{INTRODUCTION}

Pathology has played an essential role in diagnosing gastrointestinal disorders\cite{c1}. However, errors can occur due to complex systems, time constraints and variable inputs\cite{c2}\cite{c3}. This can be further complicated when the biopsy images share histological features. Computational methods have the potential to address these challenges. In light of this, developing assistive computational methods can help mitigate said errors. The goal of applying computational methods in identifying diseases is for developing fast, reproducible and reasonably accurate methods that can be easily standardized\cite{c4}.

Deep learning in detecting diseases in histopathology images has been an active area of research\cite{c5}\cite{c6}. Convolutional Neural Networks (CNN), a type of deep learning architecture, are particularly suited for distinguishing features in biopsy images. Past work in this area involves using CNN to detect cancer metastases in high resolution biopsy images\cite{c7}.

Applying CNNs to high resolution biopsy images from gastrointestinal patients may distinguish features in diseased tissues, specifically Celiac Disease\cite{c8}\cite{c9} and Environmental Enteropathy. These diseases have significantly overlapping features, which makes differentiating between the two particularly difficult\cite{c10}. 
CNNs learn from different areas of an image, look for similar patterns in new images and classify them based on feature similarity\cite{c11}. Our hypothesis is that a CNN will find differences in histologically similar tissues that are sometimes indistinguishable under a microscopic lens. We present in this paper a viable deep learning framework to classify duodenal biopsy images into Celiac Disease, Environmental Enteropathy or Normal tissues.

\section{Dataset}

Images were extracted from 465 high resolution whole slide images (WSIs) taken from 150 H\&E duodenal biopsy slides (Refer Table I). The biopsies were from children who underwent endoscopy procedures at either Aga Khan University Hospital in Pakistan (10 children \textless 2 years with growth faltering, EE diagnosed on endoscopy, n = 34 WSI), University Teaching Hospital in Zambia (16 children with severe acute malnutrition, EE diagnosed on endoscopy, n = 19 WSI), or the University of Virginia Children’s Hospital (63 children \textless 18 years old with CD, n = 236 WSI; and 61 healthy children \textless 5 years old, n = 173 WSI).

\section{Method}

\subsection{Patch Creation}

We had access to a small number of patient biopsies which were however of very high resolution and large file size. Therefore the high resolution image slides were split into patches of 1000$\times$1000 pixels and 2000$\times$2000 pixels with an overlap of 750 pixels and 1000 pixels respectively in both horizontal and vertical axes to make sure that the patches represented patterns from the slide exhaustively. We discarded patches that contained less than 50\% tissue area. The patches were then resized to 256$\times$256 pixels for feeding into the CNN model. Each slide generated an average of about 250 1000$\times$1000 patches and about 40 2000$\times$2000 patches per slide. Since Celiac Disease had the most slides and generated more patches than others, the Environmental Enteropathy and Normal patches were up-sampled by an appropriate factor to balance the data and avoid bias in the model. 

\begin{table}[htbp]
\caption{Dataset}
\begin{center}
\begin{tabular}{|c|c|c|c|}
\cline{1-3} 
\textbf{Type} & \textbf{WSIs}& \textbf{Patches} \\
\hline
Celiac& 239& 66528  \\
\hline
EE& 52& 11021  \\
\hline
Normal& 174& 41262  \\
\hline
\end{tabular}
\label{tab1}
\end{center}
\end{table}

\begin{figure*}[htbp]
\centerline{\includegraphics[scale=0.8]{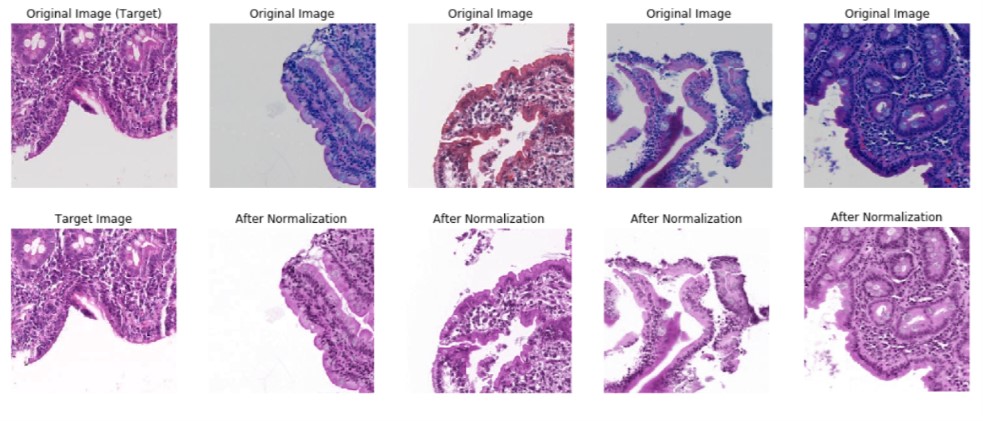}}
\caption{Stain Normalization on patches. Images in the top row highlight the difference in stain color before normalization and the bottom row shows images after the color normalization was applied using the mentioned target image.}
\label{fig}
\end{figure*}

\subsection{Stain Normalization}

There were visible variations in the color of the images due to differences in scanners, staining chemicals used while preparing the slide, and staining methods used across pathology labs. When these images were analyzed using deep learning techniques, it led to erroneous results that were not based on the features of the image but on color difference. To address this issue we applied Structure Preserving Color Normalization described by Vahadane et. al.\cite{c12} using a empirically chosen target image from the EE dataset to normalize all our images. The results are visualized in Fig. 1. 

\subsection{Image Pre-processing and Augmentation}

We performed extensive data augmentations to prevent overfitting the model. As histopathology images exhibit both rotational and axial symmetry, 4 copies of each image patch was created using a random combination of rotation (90, 180, 270 or 360 degree angle), mirroring and zoom (between 1$\times$ and 1.1$\times$). 

\subsection{Classification model}
We utilized a ResNet50 architecture to classify our patches as it has been shown to work well on computer assisted diagnosis of breast cancer\cite{c13}. To combat data sparsity we use transfer learning, which has been shown to work well on limited data, by pre-training the model on the ImageNet dataset\cite{c14}\cite{c15}. 

Since different layers capture different information, we use discriminative fine tuning as described by Howard and Ruder \cite{c16}. The layers closer to the input were more likely to have learned more general features, while the later layers identified more abstract features depending on the dataset the model has been trained on. Therefore, the learning rate used while training the initial layers was 1/9th of the rate of the final layers, while the middle layers used a learning rate 1/3rd the rate of the final layers.

Furthermore, we used cyclic cosine annealing with restarts to prevent the model getting stuck in a local minima while training\cite{c17}. By lowering the learning rate periodically, we ensured the model does not overshoot the global minima. We intermittently reset the learning rate by starting with a larger value so that it can move out of the local minima, if stuck, and reach the global optimum. By restarting, we also eliminate the need to experimentally estimate the value of our learning rate. 

Test Time Augmentation (TTA) was performed while making final predictions to ensure the predictions are insensitive to image orientation. TTA randomly performs augmentations (zoom, tilt, brightness) on the images during prediction, thus allowing the model to identify common patterns at a micro level with little regard for image orientation. The model was trained over 10 epochs with a batch size of 32. 

\begin{figure*}[ht]
\centerline{\includegraphics[scale=0.2]{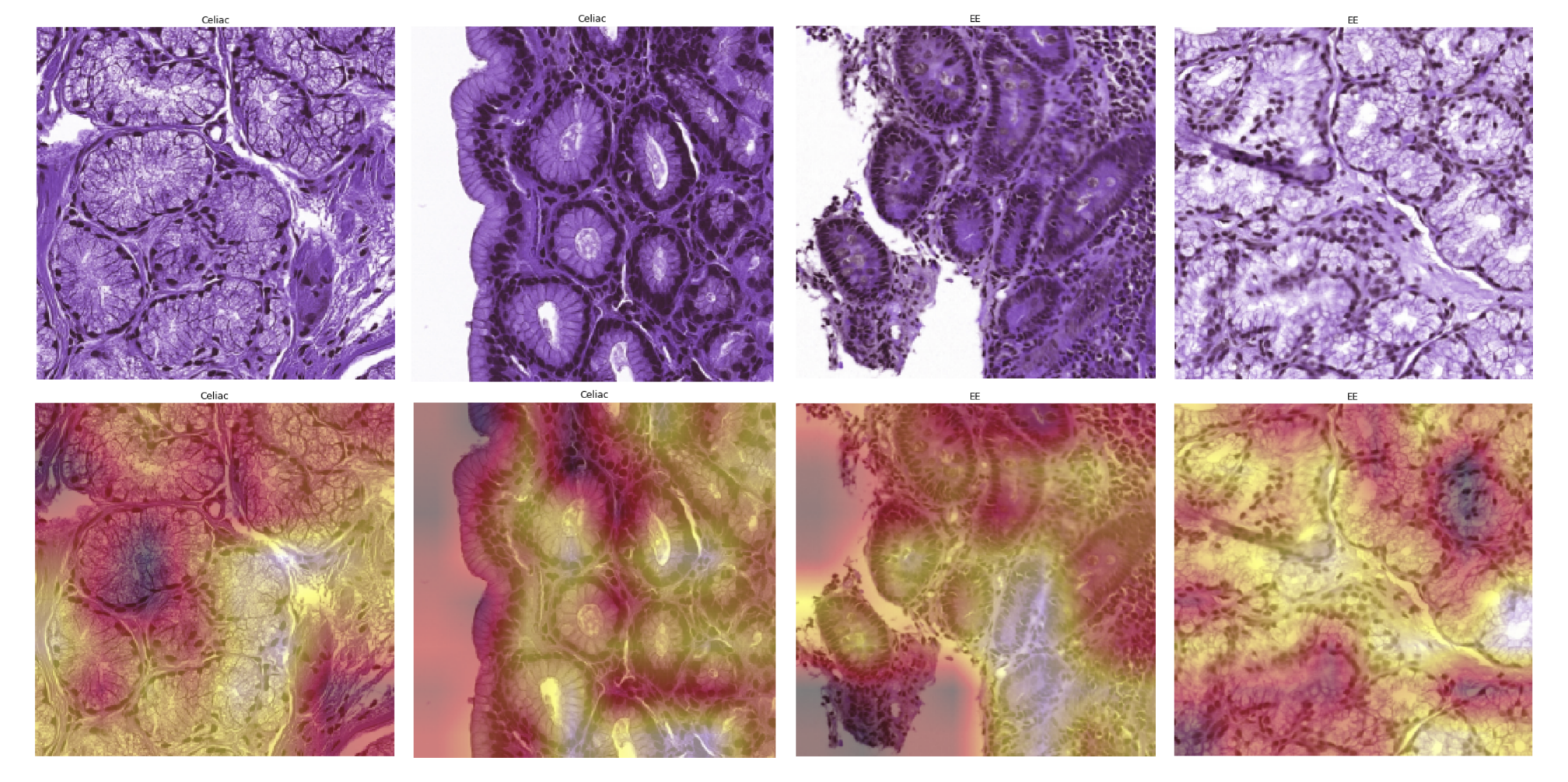}}
\caption{ Gradient-weighted Class Activation Maps for diseased tissue patches.
\textbf{Top row}: Input tissue images (\textbf{left}: two Celiac Disease patches. \textbf{right}: two Environmental Enteropathy patches.) \textbf{Bottom Row}: Corresponding heatmaps generated using Grad-CAM. Brighter areas represent higher importance.}
\label{fig}
\end{figure*}

\subsection{Multi-zoom Approach}
To create a more robust framework that looks more holistically at biopsy slides and mimics the decision making process of a pathologist, we developed a deep learning architecture made classifications based on information from multiple magnification levels of the biopsy slide. Every biopsy slide was first segmented into 2000$\times$2000 patches as highlighted and each of these patches were further segmented into patches of pixel size 1000$\times$1000 with an overlap of 750 pixels in both axes. After preprocessing of all these patches as previously outlined, color normalization was performed, after which two independent ResNet50 models were trained on these sets of 2000$\times$2000 and 1000$\times$1000 patches using the strategies described in the previous section. Additionally, all the 1000$\times$1000 patches were paired with their parent 2000$\times$2000 patches. Each pair was then passed through the corresponding trained ResNet50 model and the last fully connected layer with 2048 features from the respective models was extracted and concatenated together to give a total of 4096 features that represented the two magnification levels of a given area of the same biopsy slide. This concatenated vector gave an abstract representation of the image that was then passed through a set of trainable linear layers to make the final classification. Contrary to our expectations, we observed that this multi-scale approach provided little performance benefit weighed against the computational complexity introduced. 

\section{Results and Analysis}
\subsection{Results}
We utilized patches from 367 labelled slides for training our model which reported an overall accuracy of 88.89\% on 1000$\times$1000 patches and 86.82\% on 2000$\times$2000 patches. The predictions on the patches were then aggregated to identify the classification for their parent whole slide images. The model exhibited an overall accuracy of 92.86\% on the unlabeled 98 slides. Table II shows the metrics used to assess the performance of the model on the test set. Fig 3 shows that the model achieved exceptional certainty in identifying Environmental Enteropathy while giving an overall macro-average AUC of 0.99.

\begin{figure}[ht]
\centerline{\includegraphics[scale=0.3]{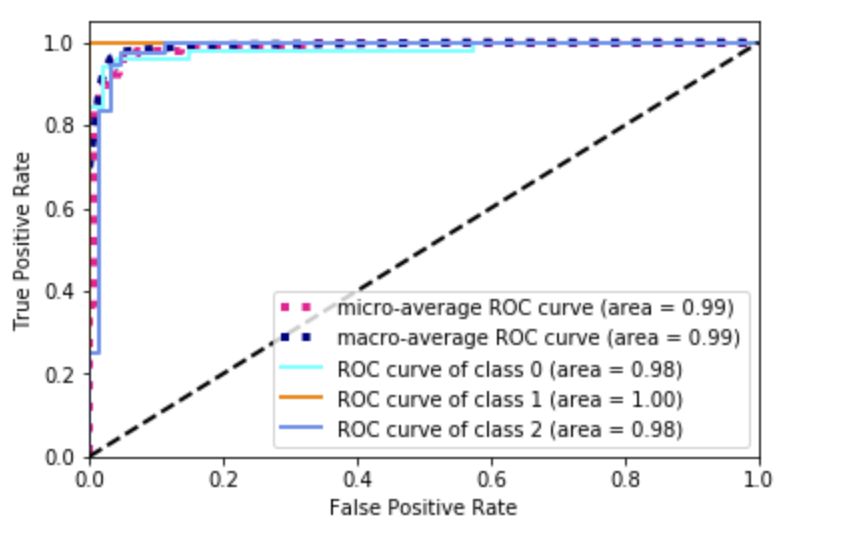}}
\caption{ROC curves on unsampled dataset}
\label{fig}
\end{figure}

\begin{figure*}[ht]
\centerline{\includegraphics[scale=0.39]{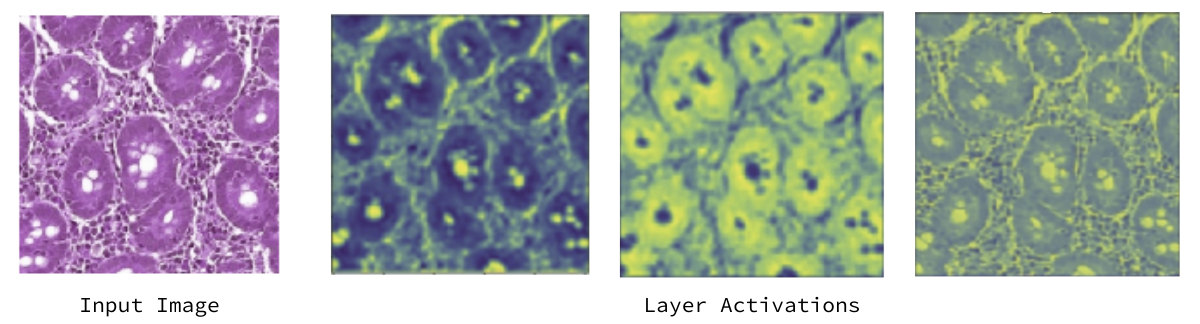}}
\caption{ Layer activation maps extracted for the given input image.}
\label{fig}
\end{figure*}

\begin{table}[htbp]
\caption{Test Set Accuracy}
\begin{center}
\begin{tabular}{|c|c|c|c|c|}
\cline{1-5} 
\textbf{Class} & \textbf{precision}& \textbf{recall} & \textbf{f1-score} & \textbf{support}\\
\hline
Celiac & 0.98 & 0.86 & 0.92 & 51  \\
\hline
EE & 1.00 & 1.00 & 1.00 & 11  \\
\hline
Normal & 0.83 & 0.97 & 0.90 & 36 \\
\hline
\textit{avg / total} & \textit{0.93} & \textit{0.92} & \textit{0.92} & \textit{98} \\
\hline
\multicolumn{5}{l}{$^{\mathrm{a}}$WSI level accuracy.}
\end{tabular}
\label{tab1}
\end{center}
\end{table}

\subsection{Interpreting the model}

It is important for us to interpret the CNN activation areas in our model for bench-marking our results with their manual counterparts. In the domain of medical imaging it is of utmost importance that we utilize methods to explain their classification result. Furthermore, visualizing activation areas allows for domain experts to corroborate the model results with incumbent classification methods. Gradient-weighted Class Activation Mapping (Grad-CAM)\cite{c18} method produces a localization map highlighting specific areas of the image that the model deems most important while making the classification decision. We implemented the methods for a residual network architecture by extracting the activation values from an intermediate convolution layer and using them to generate a heatmap identifying the important areas of the image. Viewing the images through the Grad-CAM’s lens allowed us, and domain experts, to confirm if the model was classifying the images based on real pathological features that make medical sense or image artifacts. Figure 3 displays a few example Grad-CAM outputs that were generated. 

\subsection{Visualizing Layers}
The classification was done by the CNN model by identifying patterns in the image that were associated with a certain class. Analyzing patterns that the model was looking for in the image could be useful to make sure that the image artefacts linked with a disease are being considered by the model. Therefore to understand the internal process of the designed architecture and analyze if the model was identifying all relevant image patterns, we used visualization techniques developed by Zieler et al.\cite{c19} to highlight the function of the intermediate architecture feature layers. We found that different filters in the model were successfully identifying patterns such as cell boundaries, nuclei and backgrounds in the image as shown in Fig. 4. This coupled with the Grad-Cams generated helped us conclude that the model was classifying by identifying patterns similar to what pathologists search for in the slides.





\section*{ACKNOWLEDGMENTS}

This research was supported by University of Virginia, Engineering in Medicine SEED Grant (SS\&DEB), the University of Virginia Translational Health Research Institute of Virginia (THRIV) Mentored Career Development Award (SS), and the Bill and Melinda Gates Foundation (AA;OPP1138727; SRM;OPP1144149; PK;OPP1066118).


\end{document}